\documentclass[10pt,letterpaper]{article}
\usepackage{opex3}
\usepackage[]{graphicx}
\usepackage{amssymb}
\usepackage{amsmath}
\usepackage{amsfonts}

\begin{document}

\title{Scanning plasmonic microscopy by image reconstruction from the Fourier space}

\author{Oriane Mollet, Serge Huant, and Aur\'{e}lien Drezet}
\address{Institut
N\'{e}el, CNRS and Universit\'{e} Joseph Fourier, BP166, 38042
Grenoble Cedex, France}

\email{aurelien.drezet@grenoble.cnrs.fr} 



\begin{abstract}
We demonstrate a simple scheme for high-resolution imaging of
nanoplasmonic structures that basically removes most of the
resolution limiting allowed light usually transmitted to the far
field. This is achieved by implementing a Fourier lens in a
near-field scanning optical microscope (NSOM) operating in the
leakage-radiation microscopy (LRM) mode. The method consists of
reconstructing optical images solely from the plasmonic `forbidden'
light collected in the Fourier space. It is demonstrated by using a
point-like nanodiamond-based tip that illuminates a thin gold film
patterned with a sub-wavelength annular slit. The reconstructed
image of the slit shows a spatial resolution enhanced by a factor
$\simeq 4$ compared to NSOM images acquired directly in the real
space.
\end{abstract}



\section{Introduction}
A sub-wavelength object diffracts light into evanescent and
propagating waves. It is the evanescent part - the so-called
forbidden light - that carries information on the sub-wavelength
details of the object. This evanescent contribution plays a key role
in experiments targeted at imaging surface-plasmon polaritons
(SPPs), which are electron-photon hybrid states naturally confined
at the boundary between a metal and an insulator. As such, SPPs are
strongly modified by local changes of their environment at the
nanoscale and it is therefore critical to find efficient methods to
probe the interaction of SPPs with nanostructures. This constitutes
the central motivation of the present work. In this context, imaging
SPPs can be achieved using the `forbidden light' NSOM arrangement
[1-4], where the forbidden light that couples into the sample glass
substrate at angles larger than the critical incidence
($\theta_C\simeq 43.2^\circ$ in fused silica of optical index $n
\simeq 1.46$) is collected by an elliptic mirror. An alternative
method to collect the plasmonic forbidden light uses a Fourier lens
placed in the optical path of a LRM setup to selectively image in
the Fourier space SPP beams propagating along
specific directions \cite{4}.\\
\indent In this paper, we consider a sub-wavelength slit patterned
in a thin gold film that we image with a NSOM using a point-like
optical tip for illumination. Our first motivation is to extend
previous works [1-5] by demonstrating that the use of Fourier space
SPP signals for image reconstruction of this nanostructure provides
optical images that are much better resolved (about 4 times)
compared to images acquired in the real space directly. Our second
motivation is related to the emerging field of quantum plasmonics,
where one or several quantum emitters are coupled to plasmonic
nanostructures to generate single SPP quanta [6-12]. This field is
usually dealing with weak optical signals that can be polluted by a
range of spurious incoherent light originating, e.g., from gold
fluorescence. Cleaning the useful quantum-related information from
spurious incoherent signals could be crucial for future quantum
plasmonics experiments: in addition to offering enhanced spatial
resolution, our method offers a powerful way of making this cleaning
step. This is demonstrated here by using as point-like optical tip a
scanning quantum source of light made of a single fluorescent
nanodiamond glued at the apex of a fiber tip [13-18].
\section{Experiment} \indent
A sketch of our setup is shown in Fig.~1. Basically, this is a
transmission NSOM making use of a fluorescent nanodiamond-based
optical tip \cite{12,sando} for illumination of a thin (30 nm) gold
film deposited on a fused silica cover slip. Our motivations for
using such a tip are multiple. First, the small size ($\simeq$ 25
nm) of the nanodiamond mimics a point-like source of light, which
has the potential for a better spatial resolution than standard
metal-coated optical tips \cite{13} (ultimately limited only by the
scan height of the point source over the structure). This
nanodiamond hosts three to four Nitrogen-Vacancy (NV) color-centers
as revealed
\begin{figure}[h] \centering
\includegraphics[width=12cm]{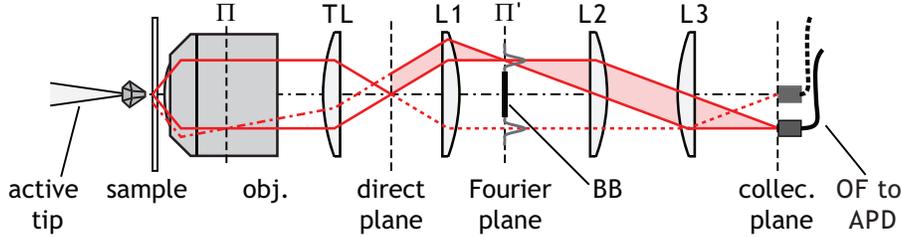}
\caption{Layout of the experimental setup: obj.= X100 oil immersion
objective of effective numerical aperture $NA = 1.35$; TL= tube
lens; $L_1$, $L_2$ (removable), $L_3$= achromatic lenses; $BB$= beam
block; $OF$= multimode optical fiber; APD= avalanche photodiode.
$\Pi'$ is the back-focal plane of $L_1$. $\Pi$ is
the objective back-focal plane and is located inside the objective
itself~\cite{4}. The $OF$-APD combination can be replaced by a
camera (not sketched) aligned with the optical axis for imaging. In
this setup, the tip is fixed and the sample is scanned in all three
dimensions with nanometer accuracy. The remaining excitation at 515
nm is removed by an optical filter (not shown). A limited number of
light rays are indicated for clarity.}
\end{figure}by time-intensity second-order
correlation measurements [15-18]. This means that the
quantum source cannot emit more than three to four photons at a time.\\
\indent The diamond is illuminated with a $\lambda_{exc.}= 515$ nm
laser light shone into the single mode optical fiber that is
terminated by the tip. As we have shown previously [16-18], the
red-orange near-field fluorescence of the NVs is able to launch SPPs
into the gold film, whereas the green laser excitation cannot do
that because of strong interband absorptions in gold in this
wavelength range: this is a second motivation for using a
nanodiamond tip. Finally, our arrangement generates a range of
undesired cross-excited nonplasmonic lights such as gold
fluorescence in addition to the useful plasmonic signal: this will
be advantageously used to demonstrate the ability of our method to
discriminate among useful and spurious signals. In the quantum
regime, where only a few photons couple to SPPs, it is crucial to
eliminate this spurious
light.\\
\indent Apart from the use of an active tip, our setup has
additional special features. First, a removable Fourier lens $L_2$
is implemented. In the absence of $L_2$, the light is collected in
the direct space in a standard way. In this case, the collection
fiber ($OF$ in Fig.~1) is optimally placed along the optical axis and the collected signal is sent to the APD. Now adding $L_2$ so
that the back-focal plane  $\Pi'$ of the microscope is conjugated
with the collection plane allows collecting light in the Fourier
space [5,16-18] (note that the $\Pi'$ plane plays
the same role as the back-focal plane of the objective $\Pi$ that
is located inside the objective itself and is difficult to access for optical imaging). In this case, the
collection fiber $OF$ can be laterally displaced to focus on the
SPP useful signal only (see below), as shown schematically in Fig.
1. In addition, an optional beam block ($BB$ in Fig.~1) of adjusted
lateral extension can be added in the Fourier plane to remove all of
the allowed light that couples into the silica substrate at smaller
incidences not exceeding $\theta_C$ [18]. Finally, the collection
fiber can be replaced by a wide-field camera to obtain either
real-space or Fourier-space instantaneous images of the SPP
propagation for a given tip position.
\begin{figure}[h] \centering
\includegraphics[width=12cm]{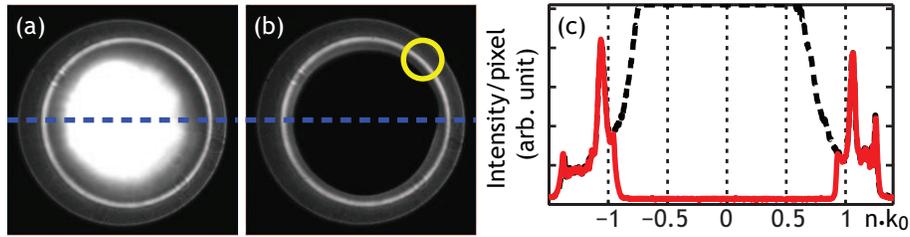}
\caption{a) Unfiltered and (b) filtered LRM images acquired in the
Fourier plane. In (b), the small circle in the upper-right part of
the SPP circle marks the imprint of the optical
fiber $OF$ used for image reconstruction. (c) Cross sections of (a) and (b)
along the blue dashed lines. }
\end{figure}
\section{Results and discussion}
\indent The principle of our method can be introduced by commenting
on the Fourier-space LRM images shown in Figs.~2(a) and 2(b), which
are obtained with the tip positioned in the near field of a 30-nm
thick gold film. The Fourier lens $L_2$ is set in and the image is
recorded on the camera. In order to filter the light
emitted in the allowed-light cone we introduce a beam block ($BB$ in
Fig.~1) centered in the Fourier space (compare Figs.~2(a) and 2(b)).
A distinctive bright circle can be seen in both images. It is
typical for a SPP field generated at the air-gold interface and
leaking into the silica substrate [2,5,16-20] at leakage radiation
angles $\theta_{\textrm{LR}} > \theta_C$, as a consequence of phase
matching at the gold-silica interface. Here $\theta_{\textrm{LR}}$
is spread over an angular $2^\circ$ range ($\theta_{\textrm{LR}}\in
[45^\circ, 47^\circ]$) because of the phonon broadening in the NV
emission used for SPP excitation \cite{21}. The SPP circle
corresponds to an effective index $n_{\textrm{SPP}} \simeq 1.06-1.07
> 1$. It therefore falls in the forbidden-light zone and, as such,
can only be detected with a $NA > 1$ objective. It represents a
clear SPP signature, essentially cleaned from any spurious light,
which couples in silica at smaller incidences up to $\theta_C$
\cite{19} as already mentioned. Cross sections shown in Fig.~2(c)
illustrate the importance of filtering the allowed light for clean
SPP imaging. From unsaturated images (not shown) we estimate the
fraction  of spurious light to $\simeq 70 \%$ of the total
integrated signal in Fourier space. Therefore, in the following, we
only focus our attention on those photons
emitted in the forbidden-light region.\\
\indent Our method consists of mapping the intensity of the SPP
circle while the plasmonic system is raster scanning under the tip.
For this purpose, the multimode fiber $OF$ is positioned in the
detection plane (Fig.~1) so as to coincide with this SPP circle.
\begin{figure}[h] \centering
\includegraphics[width=12cm]{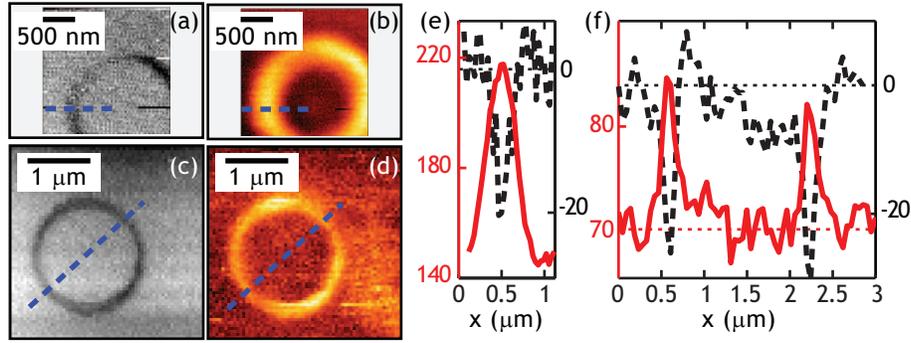}
\caption{Demonstration of the image reconstruction method. (a) and
(c) Topographic images recorded simultaneously with the optical
images (b) and (d), respectively. (b) Direct-space image obtained by
scanning the slit under the nanodiamond tip. (d) Reconstructed image
obtained by mapping the intensity of the SPP circle as function of
the slit position under the tip. (e) [respectively (f)] Cross sections
along the blue dashed lines in (a) and (b) [respectively (c) and
(d)]. Left scales stand for the optical signal levels, expressed in
units of kHz, right scales stand for the topography levels,
expressed in nm.}
\end{figure} The small circle in the upper-right part of the SPP circle in Fig.~2(b)
marks the imprint of $OF$ used for image reconstruction. Note that
only a small portion of the SPP circle will be mapped out. This does
not affect the useful signal-to-background ratio since the spurious
light is already filtered. To demonstrate our method, we have chosen
to image a circular slit of $\simeq$ 120 nm rim thickness and
$\simeq$ 1.5 $\mu$m inner diameter patterned by focused-ion beam
(FIB) milling in a 30 nm thick gold film. The results are shown in
Fig.~3. It is clear that the reconstructed image in Fig.~3(d) is
much better resolved than the direct space image depicted in
Fig.~3(b). As a matter of fact, the sharpness of the reconstructed
image competes with that of the simultaneously acquired topographic
image \cite{22} in Fig.~3(c). This is confirmed by the cross
sections shown in Figs.~3(e) and 3(f). In Fig.~3(f), the full width
at half maximum of the optical signal is around 130 nm, to compare
to 100 nm \cite{23} in the corresponding topographic cross section,
whereas in Fig.~3(e), it is as large as 230 nm, compared to 120 nm
in the topography.  Therefore, reconstructed images from the
Fourier-filtered signals made only of high spatial frequencies, i.e.
those due to SPPs leaking in the silica substrate, exhibit a four
times enhanced spatial resolution of $\simeq$ $130-100=30$ nm
compared to $\simeq$ $230-120=110$ nm obtained in the direct space.
This clearly shows the advantage of our method concerning
resolution. It is worthwhile to note that this 30 nm resolution fits well with the size of the
nanodiamond and with the typical distance between tip and surface
\cite{22} during scanning.
Therefore, our imaging method is close to reaching the optimal spatial resolution achievable with a point-like optical tip~\cite{13}.\\
\indent We now wish to give some insight into the physics governing
our observations. We first point out that after Fourier filtering
the recorded image represents a fair measure of the the local
amplitude of the excitation field at $\lambda=515$ nm. More
precisely, below the saturation regime of the NVs where we are
working, the fluorescence signal $I(\mathbf{r})$ at the tip location
$\mathbf{r}$ is proportional to the excitation rate
$R(\mathbf{r})\propto|\boldsymbol{\mu}\cdot\mathbf{E}_{\textrm{exc.}}(\mathbf{r})|^2$
that depends on the local (incident+ reflected) field at the NV
location ($\boldsymbol{\mu}$ is a transition dipole). $R$ is
strongly affected by the environment in the vicinity of the
nanostructure~\cite{bis,24}. With our detection protocol only those
emitted photons coupling directly to SPPs \emph{or} scattered by the
structure at $\theta\simeq\theta_{\textrm{LR}}$ are recorded and
contribute to the signal shown in Fig.~3(d). The increased contrast
between Figs.~3(b) and 3(d) suggests that the coupling of the
incident light at $\theta_{\textrm{LR}}$ is enhanced only in the
close vicinity of the aperture rim. This is also in qualitative
agreement with recent works [6-18] showing that plasmonic couplings
depend critically on the distance between the quantum emitter and
the nanostructure. We remark that, while the light emission with the
tip located far away from the rim is dominated by the leaky SPP
signal, the signal increase $\delta I/I\simeq 10\%$ observed
in the close vicinity of the structure results from a
competition between plasmonic and non plasmonic light (such as
scattering and spurious fluorescence) emitted at
$\theta\simeq\theta_{\textrm{LR}}$. Further analysis is required to
identify the emission channel responsible for
the increase in resolution reported in this work. \\
\indent It is worth pointing out that in Refs.~\cite{2,3b}, SPP imaging
was obtained with a classical NSOM aperture-tip illuminated at a
fixed laser wavelength. In particular, SPP fringes were observed
while scanning the tip in the vicinity of local protrusions~\cite{2}
or inside a square cavity made of grooves~\cite{3b} (see also the
theoretical analysis in Refs.~\cite{3c,2b}). The absence of fringes
in Fig.~3(d) is related to the fact that in our working regime we do not image the local photonic density-of-states at the
emission wavelengths of the NV but the local excitation variation
$R(\mathbf{r})$ at $\lambda_{ex}=515$ nm. Since no SPP is excited at this wavelength no oscillation or fringe is expected to show up~\cite{bis}.\\
\indent Before summarizing, it is worth commenting on an additional
feature of our method. It was found recently that Fourier filtering
can lead to strong image distortions in the direct space in such a
way that the intense SPP signal expected at the center of the
direct-space LRM images can be completely washed out \cite{15,16}.
This is due to interferences arising between propagating SPP modes
and the residual transmitted light (here the NV fluorescence)
\cite{17}. In the present work, we even found that it was not
possible to image the circular rim in the direct space as shown in
Fig.~3(b), but with the Fourier filter in place. This complication
is avoided by working in the Fourier space. A similar conclusion was
reached recently in experiments aimed at probing the second-order
coherence of SPPs launched by quantum emitters into a metallic film
\cite{19}. Finally, it is worth recalling that high lateral resolution imaging was
reported long time ago using the so called apertureless scanning
plasmon near-field microscope (SPNM) based on scanning tunnelling
microscopy (STM) methods~\cite{last}. It would be interesting to
compare precisely our method with this SPNM approach, in particular in view of the recent development of STM methods coupled to LRM
~\cite{rerelast,relast}.

\section{Conclusion} In summary, by probing leaky SPPs
launched by a scanning quantum Emitter, we have demonstrated a
simple method to image a plasmonic device - a circular ring in a
thin gold film - with a four times enhanced spatial resolution, i.e.
a $\simeq$ 30 nm resolution. The method implies working in the
Fourier space to reconstruct images in the direct space from SPP
signals that are cleaned from detrimental contributions with short
spatial frequencies. We believe that the approach presented here
will be of interest to applications in the emerging field of quantum
plasmonics in general, and for local studies of quantum emitters
coupled to metal films in particular~\cite{5,24}.
\\
\\
Acknowledgements: We thank G. Dantelle and T. Gacoin for providing
us with nanodiamond samples and Jean-Fran\c{c}ois Motte for the
optical tip manufacturing and FIB milling. This work was supported
by Agence Nationale de la Recherche (ANR), France, through the
PLASTIPS and NAPHO projects.

\begin{thebibliography}{99}
\bibitem{23} Slight dispersions in the measured rim width are due to an
imperfect patterning during the FIB milling process.
\end{thebibliography}
\end{document}